# Nonresonance adiabatic photon trap

S.S.Popov[1], M.G. Atluhanov[1], A.V., Burdakov[1,2], M.Yu. Ushkova[3].

[1]*Budker Institute of Nuclear Physics SB RAS*
[2]*Novosibirsk State Technical University*
[3]*Novosibirsk State University*

*Concept of high efficiency photon storage based on adiabatic confinement between concave mirrors is presented and experimentally investigated. The approach is insensitive to typical for Fabri-Perot cells requirements on quality of accumulated radiation, tolerance of resonator elements and their stability. Experiments have been carried out with the trap, which consists from opposed concave cylindrical mirrors and conjugated with them spherical mirrors. In result, high efficiency for accumulation of radiation with large angular spread and spectrum width has been confirmed. As radiation source a commercial fiber laser has been used.*

# Нерезонансная адиабатическая фотонная ловушка

С.С. Попов[1], М.Г. Атлуханов[1], А.В. Бурдаков[1,2], М.Ю. Ушкова[3]

[1]*Институт ядерной физики им. Г.И. Будкера СО РАН*
[2]*Новосибирский государственный технический университет*
[3]*Новосибирский государственный университет*

*В работе представлена и экспериментально исследована концепция высокоэффективного накопителя излучения, основанного на адиабатическом удержании фотонов между зеркалами с кривизной. Такой накопитель не ограничен типичными для ячеек Фабри-Перо условиями на качество накапливаемого излучения, точность и стабилизацию оптических элементов.*

## I. Введение

Многие фотохимические и фотокаталитические явления и их приложения занимают важное место в современной науке и технике [1][2]. К ним можно отнести и чрезвычайно важное для термоядерных установок явление как фотонейтрализация отрицательных ионов водорода и дейтерия[3]. При практическом применении воздействия излучения на атомы или молекулы желательна или необходима высокая степень насыщения поглощающих переходов. От этого существенно зависит скорость полезной реакции или выход конечного продукта, например, выделенного изотопа. Это в свою очередь требует создания устройств, формирующих необходимые потоки излучения в выделенных конечных областях пространства, где происходит та или иная фотореакция. Логично характеризовать эффективность таких устройств отношением запасаемой лучистой мощности в области реакции к отобранной от источника, например, лазера. Как правило, для решения проблемы предлагается создавать различные вариации резонансных накопителей типа ячеек Фабри-Перо [4,5,6]. Часто такие устройства или не обладают высокой эффективностью или требуют очень высокого качества излучения накачки, высокой вибро- и термостабилизации оптических элементов [7]. Действительно нужна большая добротность ячейки с одной стороны, которая, с другой, затрудняет подвод лучистой энергии. Оказывается, отказ от «резонансности» позволяет использовать излучение с достаточно большим угловым разбросом и спектральной шириной, вводимое через отверстие [8]. Удержание излучения в заданной области возможно обеспечить вогнутыми зеркалами. В работе [9] экспериментально уже исследовался накопитель из двух сферических зеркал и концепция нерезонансного удержания фотонов была подтверждена. Однако для системы [9], во-первых, характерен существенно неравномерный профиль интенсивности с большими пучностями и провалами в области удержания, а во-вторых, форма этой области мало подходит для работы с пучками частиц или быстрыми потоками газа.

Настоящая работа посвящена экспериментальному исследованию эффективности нерезонансного адиабатического накопителя фотонов с вытянутой областью занятой фотонами. Текст состоит из 5 частей включая введение. Во второй части представлена математическая модель открытого адиабатического накопителя, далее идет описание эксперимента и его результатов, следом представлены обсуждение и общее заключение.

## II. Математическая модель открытой адиабатической ловушки

Рассмотрим двумерный случай движения фотона или луча между вогнутым и плоским зеркалами Рис. 1.

Как видно из рисунка, при каждом отражении от верхнего зеркала фотон получает приращение горизонтального импульса в ту сторону, где расстояние F до нижнего зеркала больше. При малых отклонениях направления движения фотона от вертикали, он будет стремиться к центральному положению «равновесия». Найдем закон движение фотона вдоль горизонтали в представленной оптической ловушке для определения границ удержания и условий

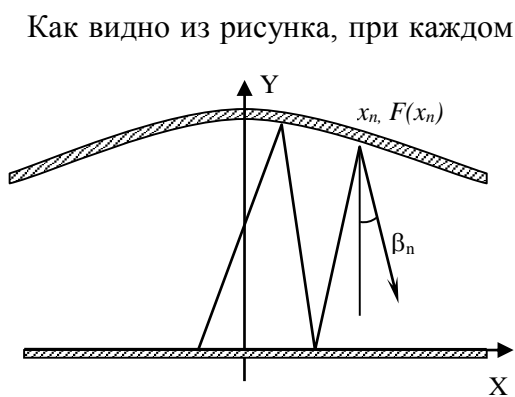

*Рис. 1. Схема адиабатической фотонной ловушки.*

устойчивости. Зададим положение фотона сразу после n-го отражения абсциссой точки отражения $x_n$, ее высотой $F(x_n)$ и углом между вертикалью и скоростью фотона $\beta_n$ (см. Рис.9). Тогда горизонтальное движение описывается следующей системой уравнений:

$$x_{n+1} - x_n = \left(F(x_{n+1}) + F(x_n)\right) tg\beta_n \qquad (1)$$

$$\beta_{n+1} - \beta_n = 2 arctg\left(\frac{dF(x_{n+1})}{dx}\right) \qquad (2)$$

Пусть $\frac{dF(x_{n+1})}{dx} \ll 1$, тогда комбинация уравнений (1), (2) позволяет выделить дискретный интеграл движения

$$\sum_n tg\beta_n (\beta_{n+1} - \beta_n) = \sum_n \frac{2(x_{n+1} - x_n)}{F(x_{n+1}) + F(x_n)} \frac{dF(x_{n+1})}{dx} \qquad (3)$$

В случае достаточной гладкости верхнего зеркала и небольших шагов, таких, что

$$\Delta F \ll F, \quad \frac{dF}{dx} \ll 1, \quad \Delta\beta \ll 1 \qquad (4)$$

интегральные суммы (3) приближенно переходят в $\ln\frac{\cos\beta_0}{\cos\beta} = \ln\frac{F(x)}{F(x_0)}$ или в стандартный адиабатический инвариант

$$F(x)\cos(\beta) = const \qquad (5)$$

Последний определяет область занятую фотонами при условии ненарастания амплитуды квазипериодического движения фотонов.

Для исследования устойчивости линеаризуем систему (1), (2), получим

$$x_{n+1} - x_n = 2F(0)\beta_n \qquad (6)$$

$$\beta_{n+1} - \beta_n = 2\frac{d^2 F(0)}{dx^2} x_{n+1}. \qquad (7)$$

Подставив одно уравнение в другое, получим следующее линейное рекуррентное соотношение:

$$x_{n+2} - 2x_{n+1} + x_n = 4F(0)\frac{d^2 F(0)}{dx^2} x_{n+1} = -4F(0)\frac{x_{n+1}}{R}, \qquad (8)$$

где $R$ – радиус кривизны верхнего зеркала. Уравнение (8) представляет тип разностной схемы для колебательной системы с единичным шагом по времени и собственной частотой $\omega_0 = 2\sqrt{\dfrac{F(0)}{R}}$. Очевидно решение представимо в виде

$$x_n = A \cdot q^n, \qquad (9)$$

где $q$ – комплексная величина. Тогда для $q$ имеем:

$$q_{1,2} = 1 - \frac{2F(0)}{R} \pm \sqrt{\left(1 - \frac{2F(0)}{R}\right)^2 - 1}. \qquad (10)$$

Условие устойчивости разностной схемы – $|q| \leq 1$, откуда, учитывая неотрицательность $\dfrac{F(0)}{R}$, получим условие «геометрического» удержания фотона

$$F(0) < R, \quad \omega_0^2 < 4. \qquad (11)$$

Приведенное ограничение совпадает с классическим условием устойчивости резонатора, образованного вогнутым сферическим плоским зеркалами, получаемым при рассмотрении бесконечной степени композиции матриц идеально фокусирующих зеркал (см., напр. [10]). Однако рекуррентный подход позволяет оценить влияние нелинейности. Действительно, при достаточно больших углах наклона лучей, когда существенно нарушается равенство $tg\beta_n \approx \beta_n$, но приращение $\beta_{n+1} - \beta_n$ остается существенно меньше амплитуды изменения угла выражение (8) примет вид

$$x_{n+2} - 2x_{n+1} + x_n = -4F(0)\frac{x_{n+1}}{\cos^2(\beta_{n+1})R}. \qquad (12)$$

Косинус в правой части обеспечивает локальное нелинейное увеличение частоты. Как видно, из Рис.8, наибольшая частота – вблизи равновесия, где фотоны имеют максимальный угол к вертикали. Тогда условие устойчивости нужно усилить

$$F(0) < R \cdot (\cos\beta_{\max})^2 \qquad (13)$$

Поправка может оказаться существенной вблизи границы устойчивости.

Как видно, радиус закругления зеркал имеет ключевое значение для удержания фотонов. Это не позволяет использовать одни плоские элементы, совмещенные под конечным углом. В последнем случае задача рассмотрена в работе [11]. В практически значимом трехмерном случае при выполнении (4) и (13) движение фотона или эволюцию луча можно очевидно рассматривать независимо вдоль осей X и Z. Кривизны верхнего зеркала вдоль этих направлений могут быть существенно разными. Несколько сложнее рассчитать случай с обоими неплоскими зеркалами, но здесь мы его подробно приводить не будем.

Представленных оценок достаточно для концептуальной разработки эффективного фотонного накопителя для нейтрализации потоков отрицательных ионов или фотохимических приложений в газовых средах. Возможная геометрия для таких приложений представлена на Рис. 2. Накопитель состоит из верхнего цилиндрического зеркала, состыкованного, например, с коническими зеркалами на концах. Радиусы цилиндра и больших оснований конусов выбраны одинаковыми. Нижнее зеркало для простоты может быть плоским. Концевые зеркала вместо конической могут иметь другую форму. Они

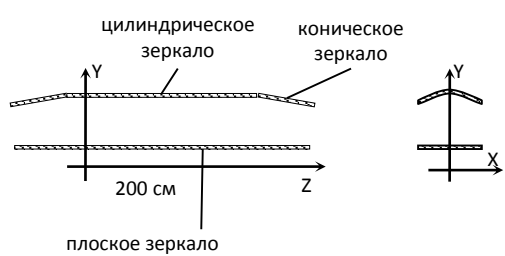

*Рис. 2 Схема квазипланарной адиабатической оптической ловушки для инжектора.*

должны обеспечивать медленное уменьшение расстояния между верхним и нижним зеркалами при движении от центра ловушки. Если относительно нижнего отразить верхние, получим эквивалентную систему из неплоских одинаковых зеркал. Z. Движение фотонов в такой системе вдоль оси Z существенно ангармоническое, а значит и образование паразитных пиков интенсивности практически невозможно. Иными словами, лучи при квазипериодическом движении вдоль Z подвержены перемешиванию, хотя и слабому. Примеры численного моделирование таких систем можно найти в [8]. Заметим, что продольное движение может оказаться неустойчивым, однако если длина накопителя достаточно велика, то время ухода фотона из ловушки превзойдет обратный декремент излучения вследствие поглощения при многократном отражение от зеркал. Если необходимо, для улучшения устойчивости в продольном направлении можно использовать плавное сопряжение центрального зеркала с концевыми. Последние в этом случае могут оказаться, например, сферическими или тороидальными. Далее влиянием расплывания ансамбля фотонов вдоль оси Z мы считаем несущественным по сравнению с поглощением при отражениях.

## Распределение плотности излучения в ловушке

Оценим плотность фотонов внутри локальной, но достаточно большой области, в которой происходит хотя бы несколько отражений. Пусть излучение инжектируется локально с малым угловым разбросом с положительной начальной скоростью и вблизи левой точки поворота. Тогда имеем уравнение непрерывности со стоком частиц в потоке

$$\langle v_z(z) \rangle n(z) = n_0 v_0 \frac{R^{m(z)} + R^{N-m(z)}}{1-R^N}, \qquad (14)$$

где нулевой индекс соответствует точке инжекции, $m(z)$ – число испытанных отражений фотонов при перемещении от источника в окрестность точки z соударений, $\langle v_z(z) \rangle \approx \langle \sqrt{c^2 - v_y^2 - v_x^2} \rangle$ усреднённая по области наблюдения продольная проекция скорости, N – число соударений на полном периоде. Второе слагаемое в числителе связано с возвращением в окрестность наблюдения после первой точки поворота. Учитывая сохранение $J = h(x,z) v_y(x,z)$, величины $m(z), N, \langle v_z(z) \rangle$ легко вычисляются

$$m(z) = \int_{z_0}^{z} \frac{J}{h(z)\sqrt{c^2 - \frac{J^2}{h^2(z)}}} dz; \quad N = \oint \frac{J}{h(z)\sqrt{c^2 - \frac{J^2}{h^2(z)}}} dz; \quad \langle v_z(z) \rangle \approx \sqrt{c^2 - \frac{J^2}{h^2(z)}} \qquad (15)$$

Если имеется несколько точек инжекции, тогда имеем ряд констант $m_i = m(z, z_{0i}, J_i); N_i = N(z_{0i}, J_i)$ и следующее распределение плотности фотонов

$$n(z) = \sum_i \frac{n_{0i} v_{0i}}{v_{zi}(z)} \frac{R^{m_i(z)} + R^{N_i - m_i(z)}}{1-R^{N_i}} \qquad (16)$$

В практически важном для нас случае, когда $1 - R^{N_i} \ll 1$ при $N_i \gg 1$ получим

$$n(z) = \sum_i \frac{n_{0i} v_{0i}}{v_{zi}(z)} \frac{2 - N_i(1-R) + O\left(((1-R)N_i)^2\right)}{N_i(1-R) - N_i(N_i-1)(1-R)^2 / 2 + O\left(((1-R)N_i)^3\right)} \approx \\ \approx \frac{2}{(1-R)} \sum_i \frac{n_{0i} v_{0i}}{v_{zi}(z) N_i} \left(1 + O\left(((1-R)N_i)^2\right)\right) \qquad (17)$$

Величина $O(\xi^n)$ такая, что $\lim_{\xi \to 0} \frac{O(\xi^n)}{\xi^n} < \infty$. Отметим, что зависимость (17) верна и в случае произвольного расположения центров инжекции для всех точек, лежащих справа от них, нужно лишь ввести эффективный источник в левой точке поворота с перенормировкой плотности тока $n_{0i}v_{0i} \to n_{0i}v_{0i}R^{-k_i}$, где $k_i$ количество отражений между левой точки остановки и действительным источником.

## III. Экспериментальное исследование эффективности удержания фотонов в открытой адиабатической ловушке

Для экспериментального исследования были изготовлен набор зеркальных элементов на подложках из монокристаллического кремния[1] с многослойным диэлектрическим покрытием. В него входили цилиндрические сегменты и несколько сферических. Длина одного сегмента составляла 50 мм, ширина отражающей поверхности около 30 мм. Общий размер зеркала мог достигать 250 мм при зоне удержания

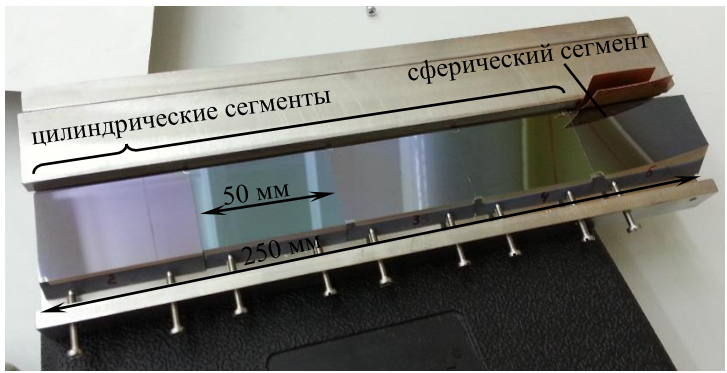

*Рис. 3. Стыковка пяти зеркальных элементов.*

150÷200 мм. Расчетный коэффициент отражения отдельных зеркал составил 0.999. Один из вариантов компоновки большого зеркала представлен на Рис. 3.

Эффективность как отношение запасаемой мощности к инжектируемой в случае устойчивого удержания очевидно определяется величиной $1/(1-R)$. В данной работе для ее измерения исследовалась зависимость динамики затухания излучения в накопителе при быстром выключении накачки от помещаемых между зеркалами различных дополнительных аттенюаторов. Также контролировался стационарный уровень накопленного излучения при постоянной накачке. Полагая, что стационарный уровень подчиняется уравнению (17), можно определить величину $1-R$ (см.(18)). Простой и наглядный примененный авторами в [9] фотометрический подход, основанный на измерении паразитного рассеяния от поверхности зеркал, для настоящей системы оказался не удобным по двум причинам. Это, во-первых, наличие большого числа стыков, дающих сильную паразитную засветку, во-вторых, большая площадь последних, что затрудняет калибровку чувствительности фотоаппаратуры по полю зрения и учет фонового излучения.

---

[1] Зеркала изготовлены в Институте лазерной физики СО РАН.

## Схема эксперимента

Схема экспериментальной установки показана Рис. 4.

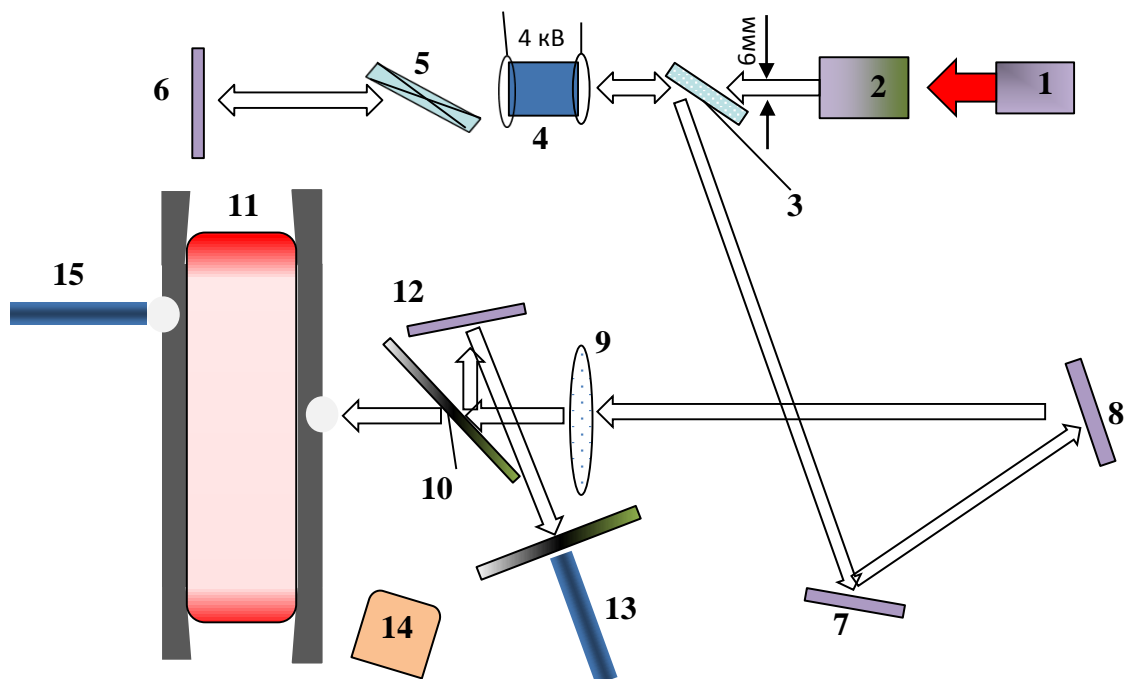

*Рис. 4. Схема измерения эффективности адиабатического накопителя. 1)лазерная головка; 2)коллиматор; 3)светоделитель; 4) ячейка Поккельса; 5) поляризатор; 6,7,8,12) зеркало; 9)линза; 10,12) светофильтры; 11) фотонный накопитель; 13,15) мониторный и сигнальный световоды; 14) CCD-камера.*

Быстрая модуляция излучения обеспечивается первыми шестью элементами. Из выходящего из лазерной головки 1 излучения специальным аттенюатором 2 формируется узкий пучок фотонов. Далее пучок пересекает стеклянный светоделитель 3, развернутый под углом Брюстера к направлению движения фотонов. После на пути стоит ячейка Поккельса 4 с поляризатором 5. Отраженное назад зеркалом 6 излучение второй раз пересекает поляризатор 5. Таким образом, возвращаемое в ячейку излучение обладает высокой степенью поляризации, и в отсутствие напряжения на ячейке излучение от светоделителя3 в направлении зеркала 7 излучение не отражается. При напряжении около 4 кВ ячейка превращает линейную поляризацию в круговую, и порядка 10% мощности попадает на зеркало 7. Устроенный таким образом затвор при снятии напряжения на ячейке позволяет достаточно быстро перекрывать поток излучения на зеркало 7 и далее. Двойное прохождение поляризатора 5 применено для более высокой степени поляризации и соответственно лучшей контрастности потока излучения на входе в ловушку при закрытом и открытом положениях. Зеркалами 7, 8 и линзой 9 излучение направляется через инфракрасный светофильтр и светоделитель 10 на входное отверстие диаметром около 300 мкм в накопителе 11. Отраженная этим светоделителем часть излучения зеркалом 12 направлялась в световод 13 для мониторирования входного излучения. Цифровая камера SDU285 [12], фиксирующая изображение входного отверстия, позволяла контролировать попадание излучения в него. Подставленный к другому отверстию световод 15 собирает небольшую долю накопленного излучения для регистрации и анализа. Динамика входного и накопленного излучений регистрировалась лавинными фотодиодами с временным разрешением порядка 2 нс. Натурная фотография оптического стенда представлена на Рис.

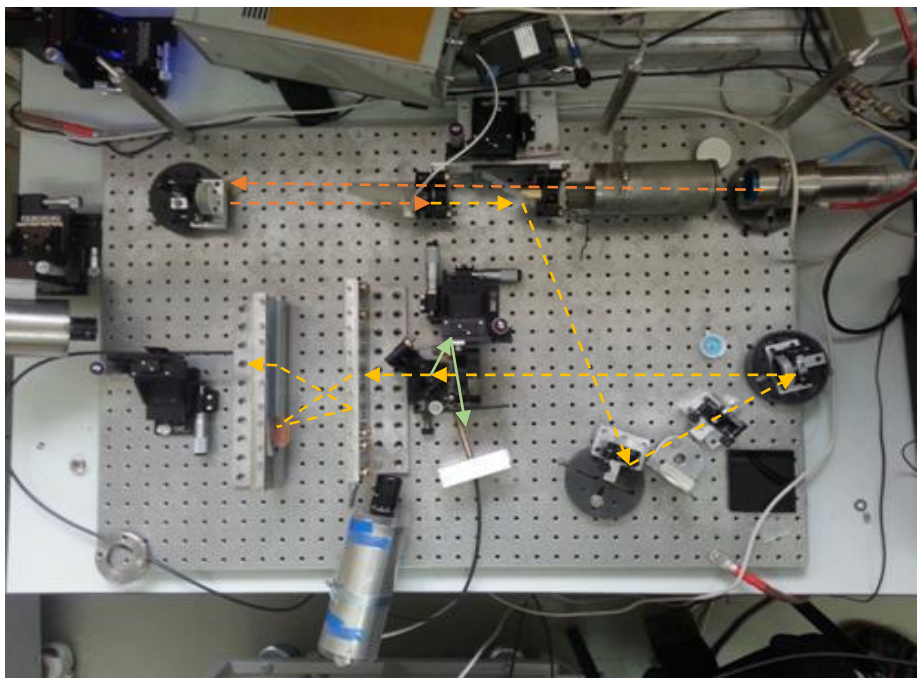

*Рис. 5. Фотография оптического стенда для исследования открытой адиабатической фотонной ловушки. Пунктирной линией показан ход луча.*

В качестве дополнительных аттенюаторов использованы решетки из вольфрамовой проволоки диаметром $d$=15 мкм (см Рис. 6). Каждый такой аттенюатор представляет собой рамку с вертикально натянутыми с определенной частотой проволочками. Коэффициент ослабления принимался равным $\alpha = \sigma d$, где $\sigma$ – частота проволочек на единицу длины.

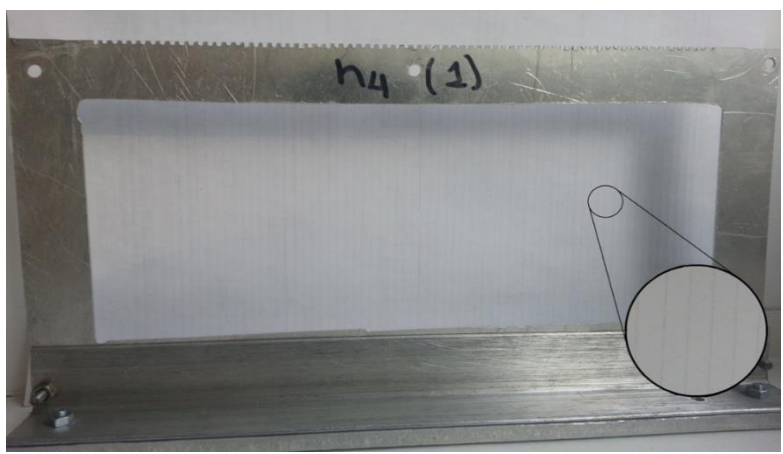

*Рис. 6. Проволочный аттенюатор.*

Источником излучения служил промышленный волоконный иттербиевый лазер [13] с длиной волны 1070 нм и шириной спектра около 7 нм.

## Измерение эффективности накопления

Осциллограммы одной из серий экспериментов приведены на Рис. 7. В этих экспериментах расстояние между зеркалами было около 12 см, угол наклона луча к оси Z – около 85º. Область удержания покрывалась пятью сегментами на каждой зеркальной поверхности с четырьмя стыками. Набор красных осциллограмм показывает динамику выключения входного излучения. Синие соответствуют динамике излучения в выходном отверстии накопителя при разном дополнительном ослаблении α. На временном интервале от 0 до 100 нс виден

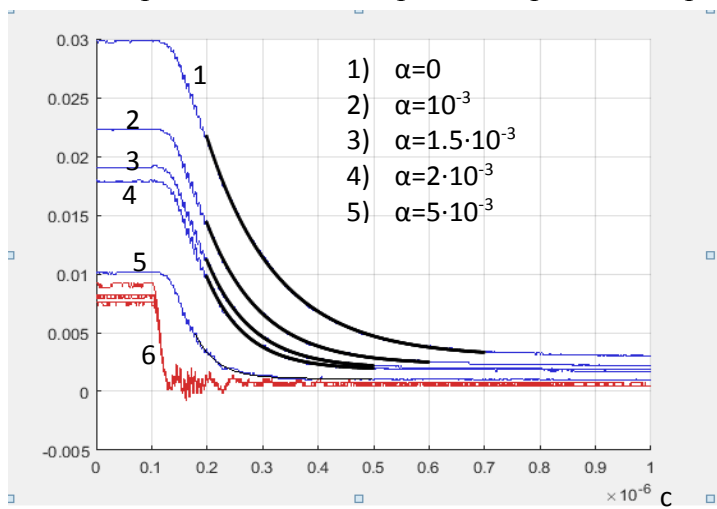

*Рис. 7. Типичные осциллограммы накопленного излучения(1-5) при разных ослабителях; жирные линии – затухающими экспоненциальные кривые; 6- осциллограммы входного излучения.*

стационарный уровень накачки накопителя. Выключение длится примерно 20÷30 нс. Затем начинается экспоненциальный спад уровня излучения, что демонстрируют аппроксимирующие черные кривые. Они достаточно хорошо согласуются с осциллограммами. Ненулевой асимптотический уровень на временах порядка 1 мкс вызван неполной контрастностью ключа входного излучения. Как видно, имеется существенная зависимость декрементов и стационарных уровней от введенного в накопитель дополнительного ослабления. Потери собственно на зеркалах можно оценить по порядку величины $10^{-3}$. Более точного измерить эффективности накопления рассмотрев соотношение стационарных уровней излучения взятых из (17) при разных потерях на одном проходе $1-(R+\alpha)$. В итоге получим

$$1-R = \frac{n(z,\alpha_k)\alpha_k - n(z,\alpha_l)n\alpha_l}{n(z,\alpha_l) - n(z,\alpha_k)} \qquad (18)$$

Аналогично можно опереться на временную динамику затухания излучения после прекращения инжекции. На масштабах времени превышающий период продольного движения *T* получим

$$n(z,t) \approx \frac{2}{(1-R)} \sum_i \frac{n_{0i} v_{0i} R^{N_i \frac{t}{T_i}}}{v_{zi}(z) N_i} \approx \frac{2}{(1-R)} \sum_i \frac{n_{0i} v_{0i} \exp\left(N_i \frac{t}{T_i}(R-1)\right)}{v_{zi}(z) N_i} \qquad (19)$$

В нашем случае с инжекцией пучка с малым угловым разбросом средняя частота отражений $\frac{N_i}{T_i}$ практически одинакова для всех запущенных лучей. Тогда из (19) имеем

$$n(z,t) \propto \exp\left(\left\langle \frac{N_i}{T_i} \right\rangle t (R-1)\right), \qquad (20)$$

где $\left\langle \frac{N_i}{T_i} \right\rangle$ -- некая характерна частота отражений для всех фотонов. Таким образом из сравнения получаемых из (20) экспоненциальных декрементов при разных ослабителях можем независимым путем вычислить эффективность накопления излучения в ловушке

$$1 - R = \frac{\tau_l \alpha_l - \tau_k \alpha_k}{\tau_k - \tau_l}. \qquad (21)$$

Хорошее совпадение установившегося экспоненциального спада на Рис. 7 подтверждает правильность (21). Оба метода измерения при сравнении осциллограмм с различными дополнительными ослабителями и без дали согласованные результаты $1 - R = (2.3 \pm 0.5) 10^{-3}$.

## IV. обсуждение

По-видимому, основной вклад в погрешность вносит нестабильность чувствительности фотодиодов, особенно, мониторного канала. Кроме этого возможно некоторое случайное влияние постановки аттенюаторов на первые проходы излучения в ловушке, когда пучок света еще недостаточно расширился. В этом случае ослабление пучка света на этих проходах может несколько варьироваться. Число таких проходов определяется расстоянием, на котором ширина пучка превысит дистанцию между соседними проволочками. В данном случае эта величина порядка 10 проходов. Так же существенными могут оказаться случайные потери на взвешенной в воздухе между зеркалами пыли. Специальных мер по ее фильтрации не предпринималось.

Очевидно негативное влияние на удержание излучения оказывают стыки между зеркальными элементами. Пример стыка показан на Рис. 8. Вероятность попасть и поглотиться на стыке в среднем за период можно оценить как

$$\mu_{sl} \sim \frac{S_{sl}}{\langle x_{n+1} - x_n \rangle} k , \qquad (22)$$

где $\langle x_{n+1} - x_n \rangle$ -- характерный интервал прицельных параметров, $S_{sl}$ -- толщина стыка, $k$ -- количество встреченных на периоде фотоном стыков. Сравним с поглощением на зеркалах за период равным $\mu \approx R^N$, учитывая, что $\langle x_{n+1} - x_n \rangle \sim 2L/N$ ($L$ – длина области удержания), получим

$$\frac{\mu_{sl}}{\mu} \sim \frac{S_{sl} k N}{2L(1-R^N)} \approx \frac{S_{sl} k}{2L(1-R)} . \qquad (23)$$

Примем общее количество стыков равным 8, $S_{sl} \approx 40$ мкм, тогда получим $\frac{\mu_{sl}}{\mu} \sim 1$. Таким образом, поглощение на стыках может оказаться существенным. Повышение эффективности накопления требует, как улучшения отражательной способности зеркал, так и уменьшения плотности стыков или их ширины.

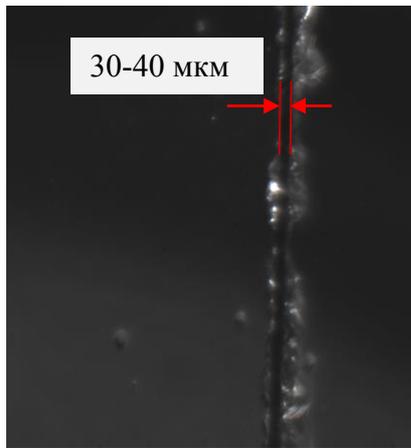

*Рис. 8. Дефекты зеркальной поверхности на стыке двух элементов.*

Значительные изменения во взаимном положении зеркал масштаба нескольких сантиметров в расстоянии и нескольких градусов в повороте не приводили к изменению эффективности накопления. Последнее доказывает нечувствительность накопителя к качеству вводимого излучения и точности выставления зеркал

V. Заключение

Традиционно предлагаемые системы для концентрирования лучистой энергии основанные на резонаторах Фабри-Перо имеют ряд сильных ограничений. Рекордные параметры, полученные, например, в [14,15], связаны с выдающимися характеристиками зеркал, малыми пространственными масштабами, и малой плотностью энергии. Достижения в создании достаточно эффективных гравитационных интерферометров, кроме прочего основаны на сложнейшей системе крепления и стабилизации оптических элементов [16,17]. Предложенный в данной работе адиабатический накопитель очевидно нечувствителен к большинству свойственных резонаторам ограничений. Он позволяет вести исследования в случаях, когда требуется высокая локализованная плотность лучистого потока и не существенно качество излучения. Это может быть многофотонная фотохимия [2], разделение изотопов и др. Особенно привлекательным такой накопитель выглядит для фотонейтрализации пучков отрицательных ионов в термоядерных приложениях[3].

Литература

1 Шпольский Э В "Современная фотохимия" *УФН* **163** (4) 87–105 (1993).

2 Макаров Г Н "Низкоэнергетические методы молекулярного лазерного разделения изотопов" *УФН* **185** 717–751 (2015)

3 J.H. Fink, A.M. Frank. *Photodetachment of electrons from negative ions in a 200 keV deuterium beam source, Lawrence Livermore Natl. Lab.* 1975, UCRL-16844.

4 Vanek V., Hursman T., Copeland D., et al., Technology of a laser resonator for the photodetachment neutralizer. Proc. 3rd Int. Symposium on Production and Neutralization of Negative Ions and Beams, Brookhaven. — 1983. — P.568-584.

5 M. Kovari, B. Crowley. *Fusion Eng. Des.* 2010, v.85 p. 745–751.

6 W. Chaibi, C. Blondel, L. Cabaret, C. Delsart, C. Drag, A. Simonin, Photoneutralization of negative ion beam for future fusion reactor. // E. Surrey, A. Simonin (Eds.), Negative Ions Beams and Sources: 1st International Symposium, AIP Conference Proceedings. V. 1097. P. 385 (2009).

7 D. Fiorucci, W. Chaibi, C. N. Man and A. Simonin. Optical Cavity Design for application in NBI systems of the future generation of Nuclear Fusion reactors. Programme and Book of Abstracts of 4th International Symposium on Negative Ions, Beams and Sources Garching, Germany 6-10 October 2014. P. 05-04

(http://www.ipp.mpg.de/3768738/programme_books_of_abstracts.pdf).

8 Popov S.S., Burdakov A.V., Ivanov A.A., Kotelnikov I.A. Photon trap for neutralization of negative ion beams. Preprint: http://arxiv.org/ftp/arxiv/papers/1504/1504.07511.pdf

9 S.S.Popov, M.G. Atluhanov, A.V., Burdakov, M.Yu. Ushkova. Experimental stady of nonresonance storage in spherical mirrors system. Optika i spektroskopiya. 2016, V. 121, № 1, p. 167–170 (in Russian.).

10 *Ананьев Ю.А.* Оптические резонаторы и лазерные пучки. М.: Наука, Гл. ред. физ.-мат. лит., 1990.264 с.

11 I. A. Kotelnikov, S. S. Popov, and M. Romé. Photon neutralizer as an example of an open billiard. Phys. Rev. E **87**, 013111, (2013).

12 СпецТелеТехника. Электронный ресурс. Режим доступа: http://www.sptt.ru/sptt/catalog.php?mod=sdu285

13 http://www.ntoire-polus.ru/HP%20fiber%20laser.pdf

14 Christina J. Hood, H. J. Kimble, and Jun Y. Characterization of high-finesse mirrors: Loss, phase shifts, and mode structure in an optical cavity. Phys. Rev. A 64, 033804 (2001).

15 G. Rempe, R. J. Thompson, H. J. Kimble, and R. Lalezari, Optics Letters, Vol. 17, Issue 5, pp. 363-365 (1992)

16 J. Aasi et al., Classical Quantum Gravity 32, 074001 (2015).

17 F. Acernese et al., Classical Quantum Gravity 32, 024001 (2015)